\documentclass[aps,prl,reprint,showpacs,superscriptaddress]{revtex4-1}
\usepackage{graphicx}
\usepackage[utf8]{inputenc}
\usepackage[english]{babel}
\usepackage{color,ulem}

\begin{document}


\title{Universal dimensional crossover of domain wall dynamics\\ in ferromagnetic films\\}


\author{W. Savero Torres}
\affiliation{Laboratoire de Physique des Solides, Universit\'e Paris-Sud, Universit\'e Paris-Saclay, CNRS, UMR8502, 91405 Orsay, France.}
\author{R. Diaz Pardo}
\affiliation{Laboratoire de Physique des Solides, Universit\'e Paris-Sud, Universit\'e Paris-Saclay, CNRS, UMR8502, 91405 Orsay, France.}
\author{S. Bustingorry}
\affiliation{CONICET, Centro At\'omico Bariloche, 8400 San Carlos de Bariloche, R\'{\i}o Negro, Argentina.}
\affiliation{Instituto de Nanociencia y Nanotecnolog\'{\i}a, CNEA-CONICET, Centro At\'omico Bariloche, R8402AGP San Carlos de Bariloche, R\'{\i}o Negro, Argentina.}
\author{A. B. Kolton}
\affiliation{CONICET and Instituto Balseiro (UNCu), Centro At\'omico Bariloche, 8400 San Carlos de Bariloche, R\'{\i}o Negro, Argentina.}
\author{A. Lema\^{\i}tre}
\affiliation{Centre de Nanosciences et de Nanotechnologies, CNRS, Univ. Paris-Sud, Universit\'e Paris-Saclay, 91120 Palaiseau, France}%

\author{V. Jeudy}
\email[]{vincent.jeudy@u-psud.fr}
\affiliation{Laboratoire de Physique des Solides, Universit\'e Paris-Sud, Universit\'e Paris-Saclay, CNRS, UMR8502, 91405 Orsay, France.}

\date{\today}

\begin{abstract}
The magnetic domain wall motion driven by a magnetic field is studied in (Ga,Mn)As and (Ga,Mn)(As,P) films of different thicknesses. In the thermally activated creep regime, a kink in the velocity curves and a jump of the roughness exponent evidence a dimensional crossover in the domain wall dynamics. The measured values of the roughness exponent $\zeta_{1d}=0.62\pm 0.02$ and $\zeta_{2d}=0.45\pm 0.04$ are compatible with theoretical predictions for the motion of elastic line ($d=1$) and surface ($d=2$) in two and three dimensional media, respectively.
\end{abstract}


\maketitle

%
%
%
%
%
Driven elastic interfaces in disordered media present intriguing scaling behaviors determined by properties such as the symmetry~\cite{Atis_prl2015}, the range of elastic interactions~\cite{ryu_natphys_2007,zapperi_PRB_1998}, the correlations of disorder~\cite{chauve_prb_2000}, and the dimensionality~\cite{Kim_Nat_2009_2D_1D,chauve_prb_2000}.
%
Experimental situations are often complex and a given dynamical system can present 
crossovers between different scaling behaviors, corresponding to different universality classes.
Some examples of this complexity are the coexistence of two distinct critical dynamics at different length scales~\cite{barabasi_stanley_1995} reported in crack propagation~\cite{santucci_epl_2010}, the crossover due to variable interaction range in ferromagnets~\cite{ryu_natphys_2007}, or to finite size effects in ferromagnetic nanowires~\cite{Kim_Nat_2009_2D_1D,kim_apex_2015_1D_2D} to name a few.
%
%
%
Investigation of universal crossovers induced by different characteristic length-scales and particularly by finite size effects are a key for understanding interface dynamics. It is also of significant technological interest for nano-devices based on domain wall (DW) motion~\cite{Parkin11042008,paruch_Nature_2016_applications}.
Evidencing a universal dimensional crossover experimentally is rather challenging and ultrathin and thin ferromagnetic films with perpendicular anisotropy~\cite{lemerle_PRL_1998_domainwall_creep} offer the opportunity to perform combined studies of interface dynamics and roughness.
In this systems, the DW creep motion is controlled by pinning energy barriers, thermal activation, interface elasticity, and the driving force, $f$, which can be due to magnetic field ($f \propto H$)~\cite{lemerle_PRL_1998_domainwall_creep,chauve_prb_2000,kolton_prb_2009_pathways}. Below the depinning threshold ($H<H_d$), the velocity follows an Arrhenius law $\ln v \sim -\Delta E/k_BT$ with an effective energy barrier decreasing with the drive $\Delta E \sim H^{-\mu}$. $k_BT$ and $\mu$ are the thermal activation energy and the so-called universal creep exponent, respectively. 
The thermal activation produces compact jumps 
(so-called thermal nuclei) allowing the interface to 
overcome pinning barriers and to advance in the direction of the drive.
The pattern of motion consists in successive avalanches 
composed of many thermally activated nuclei
with a broad size distribution~\cite{ferrero_prl_2017_spatiotemporal_patterns}.
The largest nuclei sizes, of order $L_{\mathrm{opt}}$, are 
predicted to decrease with increasing the drive~\cite{lemerle_PRL_1998_domainwall_creep,chauve_prb_2000,kolton_prb_2009_pathways} 
as $L_{\mathrm{opt}} \sim H^{-1/(2-\zeta_{eq})}$, with $\zeta_{eq}$ the 
universal roughness exponent of the DW at 
the equilibrium ($H=0$). The roughness and creep exponents are linked by the scaling relation $\mu=(d-2+2\zeta_{eq})/(2-\zeta_{eq})$, with $d$ the dimension of the interface, whose experimental verification is a stringent test of  theory~\cite{lemerle_PRL_1998_domainwall_creep,chauve_prb_2000}.

{\it Finite size effects.} Since the thermal nuclei with the largest sizes/barriers ultimately control the interface 
velocity in the creep regime~\cite{ferrero_prl_2017_spatiotemporal_patterns}, finite size 
effects are expected to occur at sufficient low drive when $L_{\mathrm{opt}}(H)$ becomes 
larger than one length-scale of 
the embedding medium.
For ultrathin films ~\cite{lemerle_PRL_1998_domainwall_creep}, the optimum length $L_{\mathrm{opt}}(H)$ always remains larger than the film thickness $t$ ($<1$~nm). The creep and roughness exponents deduced from experiments are compatible with mean field theoretical predictions ($\mu =1/4$, and $\zeta_{eq}=2/3$) for an elastic line ($d=1$) moving in a two dimensional medium ($D=2$). 
For ultrathin nanowires, the increase of $L_{\mathrm{opt}}(H)$ above the nanowire width with decreasing drive was shown to result in a dimensional crossover between the elastic line behavior ($d=1$, $D=2$) and a motion similar to particle hoping ($d=0$) along a line ($D=1$)~\cite{Kim_Nat_2009_2D_1D}.
Surprisingly, the two dimensional behavior of DWs ($d=2$, $D=3$), detected by Barkhaussen well before the development of nanotechnologies, is less well understood. Flow dynamics \cite{thevenard_prb_2011} and universal behaviors of the depinning transition~\cite{durin_prl_2000,Durin_prl_2016} have been investigated. However, to the best of our knowledge, the two dimensional behavior of DW creep motion has been evidenced yet.

In this letter, we report on evidences of two dimensional behavior of DWs creep motion and dimensional crossover due to the ferromagnetic film finite thickness. 
 We first show that below and above a crossover field the velocity curves follow the creep law $\ln v \sim H^{-\mu}$. Below the crossover field,  $\mu =1/4$ as expected for the motion of an elastic line  ($d=1$, $D=2$) and above, $\mu =1/2$ as predicted for the motion of a surface  ($d=2$, $D=3$).~\cite{lemerle_PRL_1998_domainwall_creep} A more stringent signature of the dimensional crossover is a jump of the roughness exponent $\zeta$ between two values, which are found in agreement with theoretical predictions for the quenched Edwards-Wilkinson (qEW) model with random-bond short-range pinning and including anharmonic correction to DW elastic energy~\cite{rosso_prl_2001}. 

{\it Experimental methods.}
The experiments were performed on two (Ga,Mn)(As,P) films of thicknesses $t= 12.5$ and 50~nm, and a (Ga,Mn)As 80~nm thick film, all of them presenting an out-of-plane easy magnetization axis. The films were grown by molecular beam epitaxy on a (001) GaAs substrate. The (Ga,Mn)(As,P) film was directly grown onto a GaAs buffer, while for the (Ga,Mn)As sample, a relaxed (Ga,In)As buffer was used to ensure an out-of-plane easy magnetization~\cite{lemaitre2008}. After annealing, the film Curie temperatures were 74, 130, and 126~K, respectively. 
An optical Helium flow cryostat was used to vary the temperature down to 4.3~K. DW motion was observed with a magneto-optical Kerr microscope.
The  DW displacement was induced by magnetic field pulses (duration: 1~$\mu$s to 1~s) generated by a small coil (diameter $\approx 1$~mm, rise time ~$\approx 200$~ns) directly placed on the films. 
The velocity is defined as the ratio between displacement and pulse duration. 
\begin{figure}
 \centering 
\includegraphics[width=9cm]{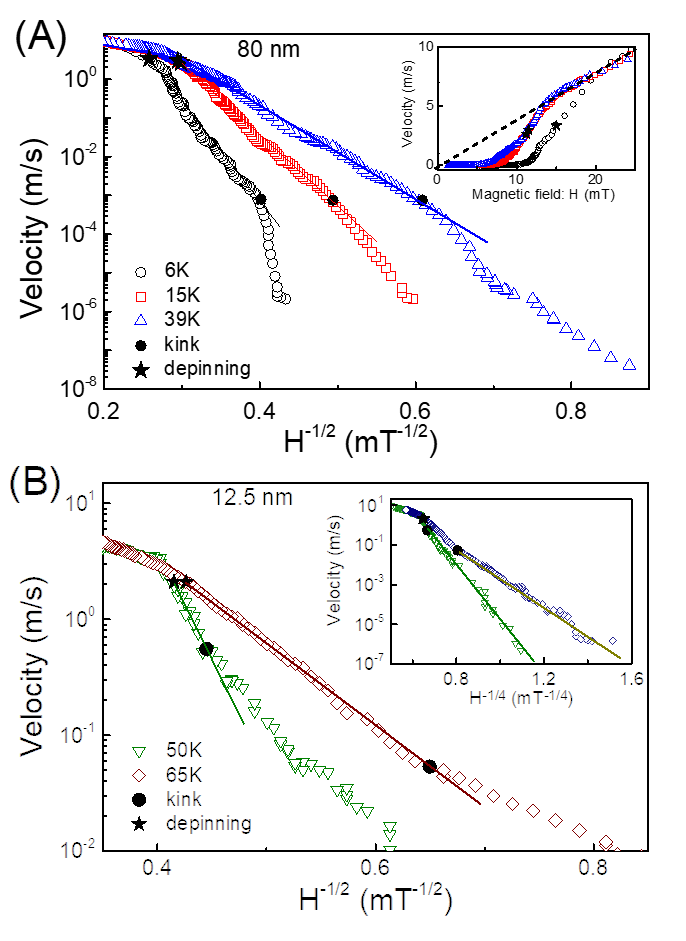}
 \caption{
 DW dynamics driven by magnetic field $H$ for (A) a 80~nm thick (Ga,Mn)As and (B) a 12.5~nm thick (Ga,Mn)(As,P) films at different temperatures. 
 The semi-log plots of the velocity versus $H^{-\mu}$ with $\mu=1/2$ (A and B) and $\mu=1/4$ (inset of B) evidence a dimensional crossover in the thermally activated creep regimes. The kinks indicated by filled circle correspond to the crossover field $H_c$. The upper limit of the creep regime $H_d$ is indicated by stars. The lines are guides for the eyes.
 The linear plot of the velocity curves (inset of A) highlights the linear flow regime (see the dashed line) observed above $H_d$.
 A sliding average over five points was performed to reduce the velocity fluctuations. 
}
\label{fig:v(H)}
\end{figure}

{\it Domain wall dynamics.}
The different regimes of DW motion driven by magnetic field are presented in Fig.~ \ref{fig:v(H)} for two different film thicknesses. 
The depinning threshold $H_d(T)$ (indicated by filled stars) corresponds to the inflection points of the velocity curves (see inset of Fig.~\ref{fig:v(H)}A). It is the thermally activated creep regime upper boundary (for details on the method see Ref.~\cite{jeudy_PRB_2018_DW_pinning}).    
For $H>H_d$, the linear, temperature independent, evolution of the velocity with $H$ (shown in the inset of Fig. \ref{fig:v(H)} A) corresponds to the dissipative flow motion. The measured mobility ($m=v/H=0.56 \pm 0.02$~m/(s.mT)) is close to the reported value~\cite{dourlat_PRB_2008} for the asymptotic so-called precessional flow regime.  

In the creep regime ($0<H<H_d(T)$), the velocity varies with temperature and magnetic field, as expected for a thermally activated motion. Surprisingly, the curves systematically display a kink (indicated by filled circles in Fig. \ref{fig:v(H)}) at a well defined crossover magnetic field $H_c(T)$. (The values of $H_c$ and $H_d$ are reported in Table \ref{table:table1}.)
For $H>H_c(T)$, the velocity curves are in agreement with the creep law ($\ln v\sim H^{-\mu}$)  for $\mu=1/2$.
For $H<H_c(T)$, the curves are compatible with $\mu=1/4$ (see the inset of Fig. \ref{fig:v(H)} B). 
As the values $\mu=1/4$ and $\mu=1/2$ are expected~\cite{lemerle_PRL_1998_domainwall_creep} for the creep motion of an elastic line and a surface, respectively, the critical exponent jump strongly suggests a dimensional crossover of the DW dynamics.

\begin{figure}
\centering
\includegraphics[width=8.8cm]{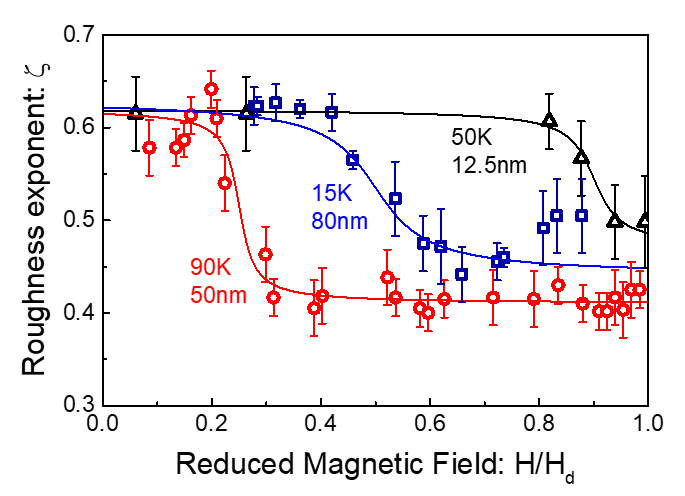}
\caption{
Roughness exponent $\zeta$ as a function of reduced magnetic field $H/H_d$ for three film thicknesses and different temperature. The steps between two constant $\zeta$-values reflect a dimensional crossover of DW dynamics. The sigmoid lines are guides for the eyes. Each error bar corresponds to the standard deviation of average value.}
\label{fig:zeta_steps}
\end{figure}

\begin{table}[h]
	\centering
	\setlength\tabcolsep{3.0pt} 
	
	\begin{tabular}{c c c c c c }
		
		\hline
		\hline
		&  & Depinning &Dynamics& Roughness  \\
		$t$~(nm) & $T$(K) &$H_d$(mT) & $H_c$(mT)& $H_c$(mT)\\
		\hline
		12 & 4.3& 8.19& 8.0(0.4)& \\
		& 10 & 7.97& 7.6(0.4)& \\
		& 30 &  7.22&6.8(0.4) & \\
		& 50 & 5.8 & 5.2(0.2) &5.7(0.4) \\
		& 65 & 5.5& 3.9(1.6)& \\
		\hline
		50  & 30 & 3.30& 2.4(0.3)& \\
		& 60 & 3.24 & 1.15(0.20) &  1.45(0.15)\\
		& 90 & 3.22 & 0.67(0.05) &  0.82(0.14)\\
		& 96 & 3.1 & 0.8(0.2)& \\
		\hline
		80 & 6   & 14.7& 6.3(0.3) &  \\
		& 15  & 11.5& 4.9(0.4) & 5.0(0.4)\\
		& 39.6& 11.15& 2.6(0.6) & 2.5(0.3)\\
		\hline
		\hline
		
	\end{tabular}
	\caption{\label{table:table1} 
	Crossover $H_c$ and depinning $H_d$ fields for the three film thicknesses ($t$) and different temperatures ($T$). The column \textit{Dynamics} (resp.~\textit{Roughness}) corresponds to the kink of velocity curves, see Fig. \ref{fig:v(H)}) (resp.~the step in roughness exponent curves, see Fig. \ref{fig:zeta_steps}). The numbers in parenthesis correspond to the crossover width.} 
\end{table}
\begin{figure}
\includegraphics[width=8.8cm]{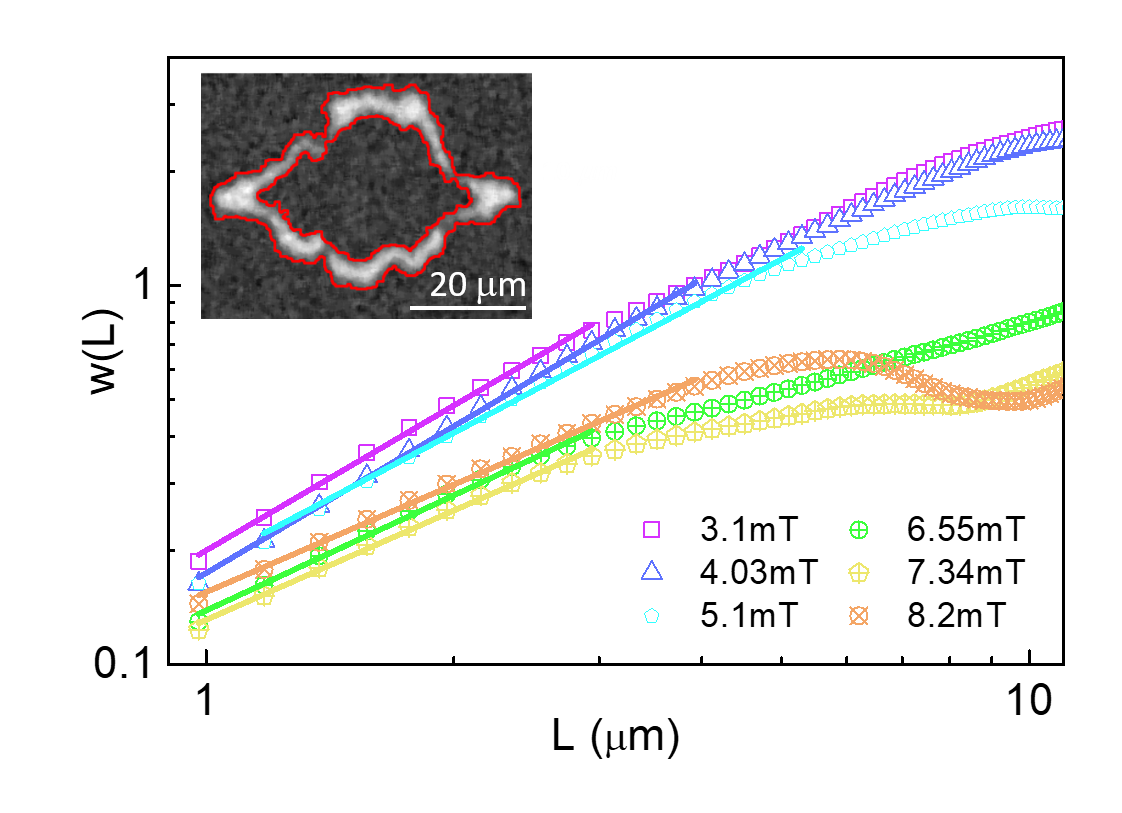}
\caption{Displacement correlation function $w$ versus distance $L$ along the DW in log-log scale obtained for different applied magnetic field values for the 80~nm thick film, at $T= 15$~K.  The lines are fits, for the small distance part, of the scaling relation $w \sim L^{2\zeta}$ (see text). Two well separated sets of curves are presenting a different slope. The slope change is observed for $H=5.1-6.55$~mT and corresponds to the step of the roughness exponent $\zeta$. 
{\it Inset.} Typical image of a DW displacement (in light gray) deduced from differential Kerr imaging showing two successive DW positions (indicated by the red contours).}
\label{fig:roughness}
\end{figure}
{\it Roughness measurements.} In order to obtain an independent and more accurate signature of the dimensional crossover, the roughness scaling properties of DWs~\cite{lemerle_PRL_1998_domainwall_creep} were analyzed in details. We calculated the correlation function of the DW displacement $u$ defined by: 
\begin{equation}
 w(L)=\left\langle \left|u(x+L)-u(x)\right|^{2}\right\rangle,
 \label{eq:roughness}
\end{equation}
where $x$ corresponds to a position on the DW, $L$ a distance from $x$ along the DW.  
In Eq.~\ref{eq:roughness}, the symbol $\langle \rangle$ corresponds to an average of measurements over all the positions $x$. As the displacement scales as $u \sim L^{\zeta}$, the displacement correlation function should follow a power law $w(L)\sim L^{2\zeta}$, with different values of $\zeta$ for the elastic line ($d=1$) and surface ($d=2$) behaviors of DW.
Experimentally, the displacements deduced from differential Kerr imaging (see the inset of Fig.~\ref{fig:roughness}) were used to determine the  displacement correlation function~\cite{lemerle_PRL_1998_domainwall_creep} as a function of $L$. Typical results are reported in Fig.~\ref{fig:roughness} in log-log scale and effectively reveal a change in the power law for $w(L)$. The distance range (1-5~$\mu$m) over which the power law changes is narrow compared to usual measurements~\cite{lemerle_PRL_1998_domainwall_creep, paruch_prb_2012_dimensional_crossover} and most probably originates from the DW displacement anisotropy (see the inset of Fig.~\ref{fig:roughness}) produced by a small in-plane magnetic anisotropy~\cite{dourlat_JAP_2007_domain_anisotropy}. 
Moreover, the curves are grouped in two sets with different slopes, at low $L$: the slope ($=2\zeta$) is higher at low field ($H<5.1$~mT) than at high field  ($H>6.55$~mT).   
In order to analyze this observation in more details, $\zeta$ was measured systematically.
%
Typical variations of $\zeta$ with reduced applied magnetic field $H/H_d$ are reported in Fig.~\ref{fig:zeta_steps}. 
%
As it can be seen, the curves present a sigmoid-like shape with upper and lower levels close to the expected values $\zeta \approx 2/3$ ($d=1$) and $\zeta \approx 0.45$ ($d=2$)~\cite{Rosso_PRE_2003}. 
Moreover, the crossover between the two levels occurs over a rather narrow range of magnetic fields ($\Delta H_c/H_d \approx 0.1-0.2$). A crossover field $H_c$ can therefore be defined. The obtained values are reported in Table~\ref{table:table1} (with the label "Roughness") for different temperature. As it can be seen, a good agreement is obtained with the values of $H_c$ (labeled "Dynamics") deduced from the kink of the velocity curves. Therefore, the step of $\zeta$-curves (see Fig.~\ref{fig:zeta_steps}) and the kink in the velocity curves (see Fig~\ref{fig:v(H)}) depict two signatures of the same DW dimensional crossover between an elastic surface ($H<H_c$) and a line ($H>H_c$).

 
%

\begin{figure}
\centering
\includegraphics[width=8.8cm]{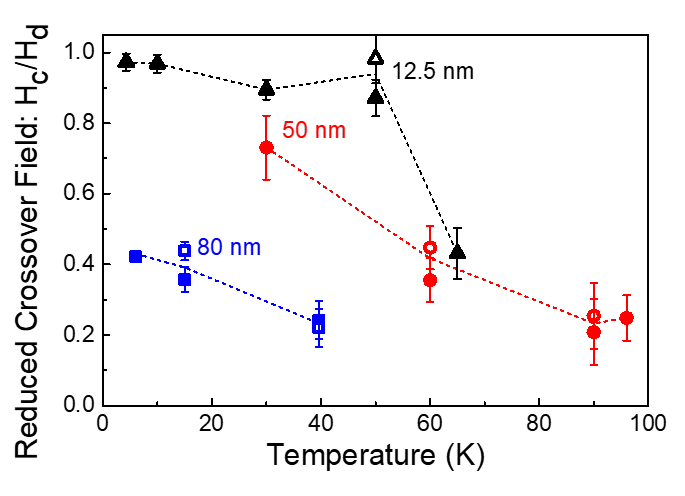}
\caption{
Temperature variation of the reduced crossover field $H_c/H_d$ for the three sample thicknesses. The filled and empty symbols correspond to the kink of velocity curves (cf. Fig.\ref{fig:v(H)}) and the step of $\zeta$-curves (cf. Fig.~\ref{fig:zeta_steps}), respectively. 
 The dimensional crossover is shifted towards the depinning transition ($H_c/H_d=1$) for decreasing film thickness. 
}
\label{fig:crossover}
\end{figure}

{\it Thermal nuclei size.} 
A summary of all the reduced crossover fields $H_c/H_d$ measured for different film thicknesses is reported in Fig.~\ref{fig:crossover}.
Interestingly, our measurements provide direct insights on  the microscopic length-scale involved in the avalanche processes producing DW motion~\cite{ferrero_prl_2017_spatiotemporal_patterns}.
Indeed, at the crossover field, the maximum excitation length 
should be close to the film thickness ($L_{\mathrm{opt}}(H_c,T) \approx t$). Therefore the film thickness ($t=12.5-80$~nm) gives an order of magnitude of the events triggering magnetization reversal avalanches. 
 %
A signature of the reduction of $L_{opt}$ with increasing drive ($L_{\mathrm{opt}} \sim H^{-1/(2-\zeta_{eq)}}$) and temperature
can be found in Fig. \ref{fig:crossover}. For example at $T=30$~K, $H_c/H_d\approx$ 0.3 (resp.~0.9) for the 80 (resp.~12.5)~nm, thick films. 
This indicates that for a fixed temperature, the range ($H_c/H_d<H/H_ d<1$) over which DW presents a two dimensional ($d=2$, $D=3$) creep motion decreases with decreasing thickness, as expected. 
 %
Moreover, for a fixed ratio $H_c/H_d$, the excitation size  $L_{\mathrm{opt}}(H,T)$ decreases with increasing temperature, which strongly suggests a decrease of avalanche sizes with increasing temperature.

{\it Analysis of the critical exponents.}
Let us now discuss the theoretical predictions and experimental results on the roughness exponent in more details. 
%
By taking into account all the measurements performed sufficiently away from the crossover, we obtain the average values $\zeta_{1d}=0.62\pm 0.02$ and $\zeta_{2d}=0.45\pm 0.04$ for $1d$ and $2d$ interfaces.
The value of $\zeta_{1d}$ is in good agreement with previous results~\cite{lemerle_PRL_1998_domainwall_creep, metaxas_PRL_07_depinning_thermal_rounding} obtained for Pt/Co/Pt ultrathin films for which the DW displacement is isotropic~\cite{metaxas_PRL_07_depinning_thermal_rounding,gorchon_PRL_2014}.
This agreement suggests that the contribution of DW displacements anisotropy (observed the inset of Fig.\ref{fig:roughness}) has a negligible contribution to the roughness. 

In ultrathin films, the  DW displacement $u(\mathbf{x},t)$ at position $\mathbf{x}$ and time $t$ driven by an external force $f$ ($\propto H$) is well described~\cite{edwards_wilkinson_1982} by the Eq.:
\begin{equation}
 \frac{\partial u(\mathbf{x},t)}{\partial t}= -\frac{\partial E_{\mathrm{el}}}{\partial u(\mathbf{x})}+f+\nu(\mathbf{x},t)+ \eta(\mathbf{x},u),
 \label{eq:qED}
\end{equation}
where $\nu (\mathbf{x},t)$  accounts for thermal noise and $\eta(\mathbf{x},u)$ for short range pinning disorder, producing DW roughness. $E_{\mathrm{el}}$ is the elastic energy which tends to flatten the DW.
In the case where only the harmonic contribution of $E_{el}$ is considered, Eq.~\ref{eq:qED} is reduced to the so-called quenched Edwards-Wilkinson equation.  
Theoretically, different roughness exponents are predicted, $\zeta_{\mathrm{eq}}$ ($f\approx 0$) close to zero drive, and  $\zeta_{\mathrm{dep}}$ ($f\approx f_\mathrm{dep}$) close to the depinning threshold. Moreover, numerical simulations indicate that $L_{\mathrm{opt}}$ is a crossover length-scale below (above) which  relevant roughness exponent is $\zeta_{\mathrm{eq}}$  ($\zeta_{\mathrm{dep}}$), respectively~\cite{kolton_prb_2009_pathways,ferrero_prl_2017_spatiotemporal_patterns}. 
Here, the spatial resolution of magneto-optical Kerr microscopy ($\approx 1$~$\mu$m) is well above  $L_{\mathrm{opt}}$ ($=12.5-80$~nm) for $H=H_c$. Increasing $L_{\mathrm{opt}}$ up to 1~$\mu$m would require to drive DW at reduced magnetic field values ($H/H_d$) lower than  $2.10^{-2}$, which is one order of magnitude lower than our experimental capabilities (see Fig. \ref{fig:crossover}).

Our experimental roughness exponents in the depinning regime can be confronted to theoretical predictions. For an elastic line ($d=1$) moving in a two dimensional medium ($D=2$), analytical calculations~\cite{chauve_prb_2000} and numerical simulation~\cite{rosso_prl_2001,Rosso_PRE_2003} predict $\zeta_{\mathrm{dep}}=1.25$ for harmonic ($\sim u^2$) and $\zeta_{\mathrm{dep}}=0.635 \pm 0.005$ for anharmonic variations of $E_{\mathrm{el}}$. Only the latter prediction is compatible with our experiments ($\zeta_{1d}=0.62\pm 0.02$). Note also that the predicted depinning and equilibrium ($\zeta_{\mathrm{eq}}=2/3$) values are too close to be discriminated experimentally.  
%
%
For the motion of elastic surface ($d=2$) in a three dimensional medium ($D=3$), the predictions are $\zeta_{\mathrm{dep}}=0.75$  for harmonic and  $\zeta_{\mathrm{dep}}=0.45$ for anharmonic variations of $E_{el}$, respectively~\cite{rosso_prl_2001,Rosso_PRE_2003}. Here also, our experimental results ($\zeta_{2d}=0.45\pm 0.04$) are only compatible with predictions assuming anharmonic elasticity for the depinning and equilibrium values ($\zeta_{\mathrm{eq}}=0.41$)~\cite{Rosso_PRE_2003}.

In conclusion, we have evidenced two concomitant 
signatures of a dimensional crossover in the thermally activated creep motion of magnetic DWs. 
On the theoretical front, it would be particularly interesting to investigate the dimensional crossover and in particular the contribution of anisotropy between the vertical and in-plane direction due to perpendicular anisotropy~\cite{thevenard_prb_2011}, which suggests the formation of strongly anisotropic thermal nuclei.
Moreover, the DW universal behavior as an elastic surface should be encountered close to the depinning threshold in any films or multi-layers with thicknesses larger than a few tens of nanometers~\cite{jeudy_PRB_2018_DW_pinning}.
As the motion can be described by a single minimal model ignoring the magnetic structure of DWs, the dimensional crossover should present a universal character and be encountered in other systems than magnets.


\begin{acknowledgments}
We wish to thank L. Thevenard and C. Gourdon for the loan of the $80 nm$ thick sample.
S. B., and V. J. acknowledge support by the French-Argentina project ECOS-Sud No.~A12E03. This work was also partly supported by the french projects DIM CNano IdF (Region Ile-de-France), the Labex NanoSaclay, No. ANR-10-LABX-0035 and the French RENATECH network. R.D.P. thanks the Mexican council CONACyT for the PhD fellowship No.~449563. S. B. and A. B. K. acknowledge partial support from Project PIP11220120100250CO (CONICET).
\end{acknowledgments}


\bibliography{refs_para_D,refs_2D_3D_crossover}


\end{document}